04.02.05

# Novel Single Photon Detectors for UV Imaging

P. Fonte[1], T. Francke[2], N. Pavlopoulos[3], V. Peskov[3], I. Rodionov[4]

[1]ISEC and LIP, Coimbra, Portugal,
[2]XCounter AB, Danderyd, Sweden,
[3]Pole University Leonard de Vinci, Paris
[4]Reagent Research Center, Moscow

## Abstract

There are several applications which require high position resolution UV imaging. For these applications we have developed and successfully tested a new version of a 2D UV single photon imaging detector based on a microgap RPC. The main features of such a detectors is the high position resolution - 30 µm in digital form and the high quantum efficiency (1-8% in the spectral interval of 220-140 nm). Additionally, they are spark- protected and can operate without any feedback problems at high gains, close to a streamer mode. In attempts to extend the sensitivity of RPCs to longer wavelengths we have successfully tested the operation of the first sealed parallel-plate gaseous detectors with CsTe photocathodes.
Finally, the comparison with other types of photosensitive detectors is given and possible fields of applications are identified.

## 1. Introduction

There are several applications which require high position resolution UV imaging. Examples include: RICH, spectroscopy, hyperspectroscopy and various security checking devices. For these applications we have recently developed 1D and 2D VUV (140-220nm) imaging detectors based on microgap RPCs [1]. Their spectral sensitivity was determined by the photosensitive elements used for this purpose: either gases with a low ionization potential or a CsI photocathode. The counting rate characteristics of RPC are determined by the resisitivity of the materials used for manufacturing their anode and cathode planes. The cathode of the detectors mentioned above was made of low resistivity materials ($\rho<10^8\Omega$cm) and was thus able to operate at extremely high counting rate N up to ~$10^5$Hz/mm$^2$ [1]. If one exceeds some critical gain (which depends on a value of N) then a glow discharge may appear [2]. It does no harm the front- end electronics or the readout strips. However, glow discharges are very sporadic, could be repeated in short periods of time and thus do not allow in an easy way to achieve a high time resolution.
The aim of this work is to investigate whether the "timing RPCs" [3] could be used as photosensitive detectors. Timing RPCs are made of highly resistivity materials ( $\rho>10^{10}\Omega$cm). In these detectors at high gains avalanches transit to streamers rather than to glow discharges. The high gain avalanche mode operation with presence of some streamers allows to achieve very good time resolutions: better then 0,1 ns. The high resistivity of materials restricts of course the rate characteristics of the detector; however, they still remain perfect for such applications as RICH, time if flight $BaF_2$ PET [4], 3D VUV locators [5] (see paragraph 4). The other aim of this work is to extend the sensitivity of gaseous detectors to longer wavelengths by using CsTe photocathodes. Note that it was demonstrated earlier that parallel-late type gaseous detectors combined with photocathodes sensitive to visible light (SbCs, GaAs and others) could operate in principle, however the maximum achievable gain is <100 [6]. This should be compared with gains $>10^5$ achieved easily with CsI photocathodes [7]. The reason why the gain of gaseous detectors combined with SbCs photocathodes is so low is well understood and described in [8] (see p.3). In the past we succeeded to boost the maximum achievable gain of gaseous detectors combined with SbCs photocathodes by exploring cascaded hole-type structure, for example capillary plates [9]. Such structures allowed one to efficiently suppress photon feedback and also reduce ion feed backs [10,11]. Coming from the results presented in [9] one can expect that with CsTe photocathodes one may reach high gains even with a simple parallel- plate structure without using the cascaded hole type structures. Indeed, the probability of the photon feedback $\gamma_{ph}$ and ion feedback $\gamma_+$ are:

$\gamma_{ph} = \int I\,(A, E, \lambda)\; Q\,(\lambda)\; d\lambda$

and

$\gamma_+ = k(\varepsilon - 2\varphi)$, where

I is avalanche emission spectra,
A is a gas gain,
E-the strength of the electric field,
$\lambda$-wavelength,
Q-the cathodes quantum efficiency,
k-coefficient,
$\varepsilon$-an ionization potential of the gas,
$\varphi$- photocathode affinity (work function).

Since the CsTe photocathode has the function $Q(\lambda)$ and the value of $\varphi$ being in between the corresponding functions and values of $Q(\lambda)$ and $\varphi$ for the CsI and the SbCs photocathodes, one can expect the maximum achievable gain $A_{max}$ will be also between the maximum achievable gains of the detectors with CsI and SbCs photocathodes. This expectation was supported by preliminary experiments with a sealed single-wire gaseous detector with CsTe photocathodes [12]. Gains close to $10^4$ were achieved with this detector. However , strictly speaking the avalanche light emission in parallel- plate detector geometry and cylindrical geometry could be different since the strength of the electric filed is different [13]. As a result the feed back probability and the maximum achievable gains could be different. Among various possible parallel- plate structures designs a special place belongs to RPCs. This will tremendously simplify the detector's design and at the same time allow it to be spark-protected. Thus one of the aims of our work was to verify this experimentally. Unfortunately, the sensitivity of the CsTe photocathode could be easily affected by tiny impurities and this is why such detectors should be made of carefully chosen materials and be carefully sealed. The long-term stability of such sealed detectors could be achieved only in low counting rate applications because at low rates the outgassing and aging are small. Thus low rate RPCs could be one of the best options.

## 2. Experimental Set Up and RPC Designs

An experimental set up for studies of the RPC is shown schematically in Fig 1. It contains a VUV source (an Hg or a pulsed $H_2$ lamp [14]), and a gas chamber inside which an RPC was installed. In most measurements the gas chamber was attached to an alignment table allowing controllable movement in 2D direction to take place with a micron accuracy, as well as rotation with the accuracy of a minute. A collimator with narrow- band filters or a VUV spectrograph could be installed between the UV lamp and the gas chamber. The later was used for detectors quantum efficiency measurements (see [1] for more details). The gas chamber was flushed by He+0.8%$CH_4$+EF [1] or $C_2H_2F_4$+10%$SF_6$+5%isobutene at a total pressure of 1 atm.

Two types of detectors were tested in this work: modified "timing RPCs"[3] and gas filled photodiodes with CsTe photocathodes.

The modified "timing RPC" developed in the frame of the present work is shown in Fig. 2. Its cathode was made of a $MgF_2$ plate 40x40mm$^2$ and 4mm thick. The inner surface of the $MgF_2$ plate facing the anode was coated by a 20 nm thick CsI layer. No metallic coating was used to ensure high surface resisitivity of the cathode. The outer surface of the $MgF_2$ plate was coated by a 0,2 μm thick Al mesh manufactured from microelectronic technology. It had a pitch of square cells of 30 μm with the width of metallic part of 7 μm. A cathode design was also tested having only metallized strips of 50 μm pitch. The anode of the RPC was made either of Pestov glass coated by Cr strips of a 30 μm pitch or of a G10 plate with gold coated pixels of 1x1 mm$^2$. The gap between the anode and the cathode was 0,4 mm. The design with strips on the outer surface of the $MgF_2$ cathode or with pixels on the G10 anode was oriented on 2D measurements. Anode strips were connected individually to an ASIC readout chip. In some simple measurements 20 central strips were connected to charge sensitive amplifiers and other strips were grounded through a 3MΩ resistor. In the case of the pixel readout 9 of the central pixel were connected to charge sensitive amplifiers and others were grounded through a 3MΩ resistor.

The ID and 2D position resolutions of this RPC were measured using collimators providing either a cylindrical beam of 30 μm in diameter or a rectangular beam 5 x 0,03 mm$^2$.

Prototypes of sealed gaseous detectors with CsTe photocathodes were manufactured by us from commercial vacuum photodiodes: Hamamatsu R702 and R1107, or from custom –made photodiodes ordered from the Photonic Company MELZ, Moscow. One of the MELZ photodiodes was manufactured by a special way: without metalisation of the window on which the semitransparent CsTe photocathde was evaporated. These photodiodes were "converted" to gaseous detectors by the following procedure: they were installed inside a special T-shaped sealed gas chamber having a $CaF_2$ window, HV feed-trougths and mechanical manipulators (see [15] for more details). The chamber was pumped for a few days to a vacuum better than $10^{-6}$ Torr after which either a He+0.8%$CH_4$+ EF or an Ar+10%$CH_4$, or an Ar+5%isobutene gas mixture was introduced. At this moment the small glass sealing in the photodiode envelope was broken by a manipulator so that the gas could entered the photodiodes. After filling with gas the MELZ photodiodes became in fact "RPCs" because these detectors had a metallic anode and a cathode made of an insolating material. Of course "classical" RPCs like the original Pestov one [16] or" timing" RPCs [3], have opposite structure: metallic cathodes and dielectric anodes, however RPCs can operate if the anode is metallic and the cathode is dielectric. In all tests the HV feeding of gas -filled photodiodes was done via 3MΩ resistors.

The gain of the detectors was measure using a pulsed $H_2$ lamp, allowing one to obtain a clear signal even at a gain of one. We also performed some stability and aging measurements using the set up described in [17].

## 3. Results

Fig. 3 a,b shows the results of position resolution measurements at two gains in the He+0.8% $CH_4$+EF mixture. In these measurements only one central strip (number 10 in the given plot) on the anode plate was illuminated by a well- collimated UV beam (30 μm in diameter) and a counting rates from this and several neighboring strips were measured simultaneously. One can see that in the He-based mixture at gains $<10^5$ the counting rate vs. the strip number has a sharp maximum at the strip which was illuminated by UV (Fig. 3a). Thus a position resolution of 30 μm was achieved at low gains. However, with further increases of the gain the counting rate profile begins to smear due to feedback problems and the position resolution degrades (Fig. 3b). Fig. 4 shows image of two microslits with width of 250 μm and 150 μm. In this particular case strips were readout by ASIC.
In the $C_2H_2F_4$+10%$SF_6$+5%isobutene mixture the position resolution remained good until a very high gain of up to $10^6$ (Fig. 5). However, in this mixture some degradation of the quantum efficiency with the time was observed. In this mixture the counting rate vs. the voltage curve had some kind of "plateau" region which is very important in single photon counting Fig. 6. Signal amplitudes vs. rate measured at two gas gains $1,5\times10^6$ and $5\times10^6$ are presented in Fig. 7.
One can see from these data that at gains of $\sim10^7$ the RPC could operated at counting rates of 100Hz/cm$^2$. This value is rather close to the rate characteristics of the conventional muon RPCs (with $\rho<10^{11}\Omega$cm) used for tracking.
A detector with a pixelized G10 anode allowed one to obtain 2D images with a position resolution of 1x1mm$^2$ in digital form-see Fig. 8.
Fig. 9 shows the quantum efficiency vs. the wavelength for the RPCs, vacuum photodiodes and gas- filled photodetectors. One can see that the quantum efficiency of semitransparent CsI photocathode was sufficiently high in order to be used in RICH or other low UV flux measurements.
In the case of the photodiodes with CsTe photocathodes their quantum efficiency dropped after the gas filling, but then remained stable for a long time ( Fig.10) The crucial point of course is to realize if the gas- filled photodiodes used in this work can operate at some gain. As we have mentioned in the introduction, single wire detectors with CsTe photocathodes could operate at gains of $10^4$. However, it is not evident that the same gain could be achieved with gas- filled photodetectors, because: a)their design is not optimized for high gain operations and b)as was mentioned above, the emission spectra of Townsend avalanches in parallel-plate detector's geometry ( medium electric fields) could be different form the emission spectra of avalanches in the cylindrical geometry (very strong electric field near the wire).
We have discovered that Hamamasu gas- filled photodiodes could operate at gains of 6-10 and that their maximum achievable gain was limited by sparks on sharp edges of the metallic anode plate. Due to cathode resistivity, these sparks were weak and did not damage the detector. One could reach gains of 100 with a UV beam well collimated to center of the detector (to avoid edge illumination) and at very low UV fluxes . MELZ photodiodes had rounded edges and as a result could operates at gains up to $10^3$ with a low intensity collimated beam (Fig.11). Some preliminary results of aging measurements are presented in Figs 12 and 13.

## 4. Comparison to other detectors and new filed of application of photosensitive RPCs

It will be interesting to compare developed RPCs with other types of photosensitive detectors. Nowadays, the most popular photosensitive gaseous detectors are wire chambers [18] combined with CsI photocathodes and cascaded capillary plate [19] or GEMs combined with CsI photocathodes [20]. Compared to wire chambers the photosensitive RPCs have better time and 2D position resolutions, can operate at higher gains and are spark-protective. Compared to capillary plates and GEMs, photosensitive RPCs have a much simpler design, in other words, a compact planar geometry. Compared to solid –state detectors photosensitive RPCs have practically the same position resolutions, but they much cheaper and do not require any cooling (see a review paper [21]). Among vacuum detectors a serious competitor to the photosensitive RPCs are only advanced MCPs having a position resolution of 12 μm [22]. However, MCPs are limited in size(they are less than 10x10cm$^2$). Photosensitive RPCs operating at 1 atm have no mechanical constrains on the window size and thus could be made at very large area. Thus our RPCs can compete with MCPs in applications which require large sensitive area, for example RICH, 3D locators [5], hyperspectroscopy [23]. The applications of photosensitive gaseous detectors in RICH are described in many papers and conference proceedings (see for example [18] and references there in).

For this reason in this report we would like to focus on much less known applications requiring large sensitive area such as 3D UV locator and hyperspectroscopy. 3D UV locator is based on a pulsed UV source (few tens of ps) and an optical system combined with position sensitive UV detectors. By a simultaneous measurements of the position of the detected photon in the focal plane of the optical system and the time delay between the UV pulse and detection of the particular photon one can reconstruct the with a high position resolution a 3D image of the investigated object.

Hyperspectroscopy is a new power method of remote surface image taking, providing simultaneously high position and spectral resolution.

The spectral resolution of this method is around 1nm, i.e. hundred times better than colour pictures. This allows to conclude from the analysis of the reflected light about the chemical composition of the surface. As an example the Fig. 14 shows two superimposed images: a usual photo (grey) and a hyperspectrographical synthesis image (colored) [24]. In the gray image one can only recognize some structures on the surface. In contrast, from the hypersepctrographic computer synthesized image one can identify the surface composition, e.g. the yellow color is sand, the rouse color wheat earth, the blue color represents swam. One can clearly see from this figure the advantages of the hyperspectroscopy over usual image taking. Up to now the hypersepctroscopic method is used in a passive mode when the reflective by the surface sun light is detected by a sophisticated optical system combined with a special type of spectrograph readout by advanced MCPs [22] having position and spectral resolution.

However hypersepctroscopy could be enforced by an "active" mode for example by using an external source with selected emission spectra, for example a UV sources (see Fig. 15). Large area photosensitive RPCs can allow one to obtain complimentary images in the UV region of spectra. Indeed, the reflection coefficient of the UV light is very small and one has to deal with very small UV fluxes. Thus large area optical system and large area detector are absolutely necessary. In this particular application photosensitive RPC will be superior vacuum MCPs.

Gaseous detectors with CsTe photocathodes may also have a lot of applications as for example in a multilayer time of flight $BaF_2$ PET [4] or flame imaging [25].

The concept of the multilayer time if flight $BaF_2$ PET is presented in Fig.16 The multilayer design allows one to reduce the parallax error in the image reconstruction. Preliminary tests show that with the compact planar detectors with CsTe photocathodes one to obtain from few hundreds to 1000 primary photolectons per Mev of the deposit energy in the $BaF_2$ crystal, sufficient to achieve good energy and time resolutions. For this application the ideal of course will be to use timing photosensitive RPCs

The concept of position sensitive fire detector is presented in Fig. 17. It contains a UV position sensitive detector with CsI or CsTe photocathodes combined with an optical system.

It is known that flames in air emit in UV region [26]. Such detector allows to obtain one UV images in the focal plane of the optical system and thus to determine the position of flames in space. Imaging capability makes this system multifunctional: it could not only obtain images of flames, but also monitor and if necessary reject UV background or confusing UV sources like sun. This detector could also be combined with several modulated external UV sources, which give controllable monitoring signals in corresponding channels of the position-sensitive detector. Appearance of smoke or mist will attenuate this signal and allows one not only detect their appearance, but also visualize them in space.

## 5. Discussion and Conclusions

We have developed a photosensitive RPC with a semitransparent CsI photocathode able to detect single photons with wavelengths of $\lambda<220$nm. The main features are: extremely high position resolution (30μm in digital form), sufficiently high quantum efficiency and compact planar geometry which s important for some applications, as for example, for RICH. They also are spark- protected and thus are able to operate at high gains, necessary to achieve good timing properties [27].
In addition we have tested gas -filled photodetectors with CsTe photocathodes. These detectors were sensitive to wavelengths up to 300 nm. Some designs of these detectors were able to operate at gains of $A_{max}=10^3$ and they were also spark -protected due to the high resistivity of their cathodes. The achieved gain is in between of $A_{max}$ achieved with CsI and SbSc photocathodes-see Fig. 18. The successful tests of these detectors opens the possibility to build sealed RPCs with CsTe photocathoes which cold find many applications, as for example in RICH, spectroscopy [1], hyperspectroscopy [23], time of flight $BaF_2$ PET[4], UV locators [5], flame detectors [25].

**Figures**

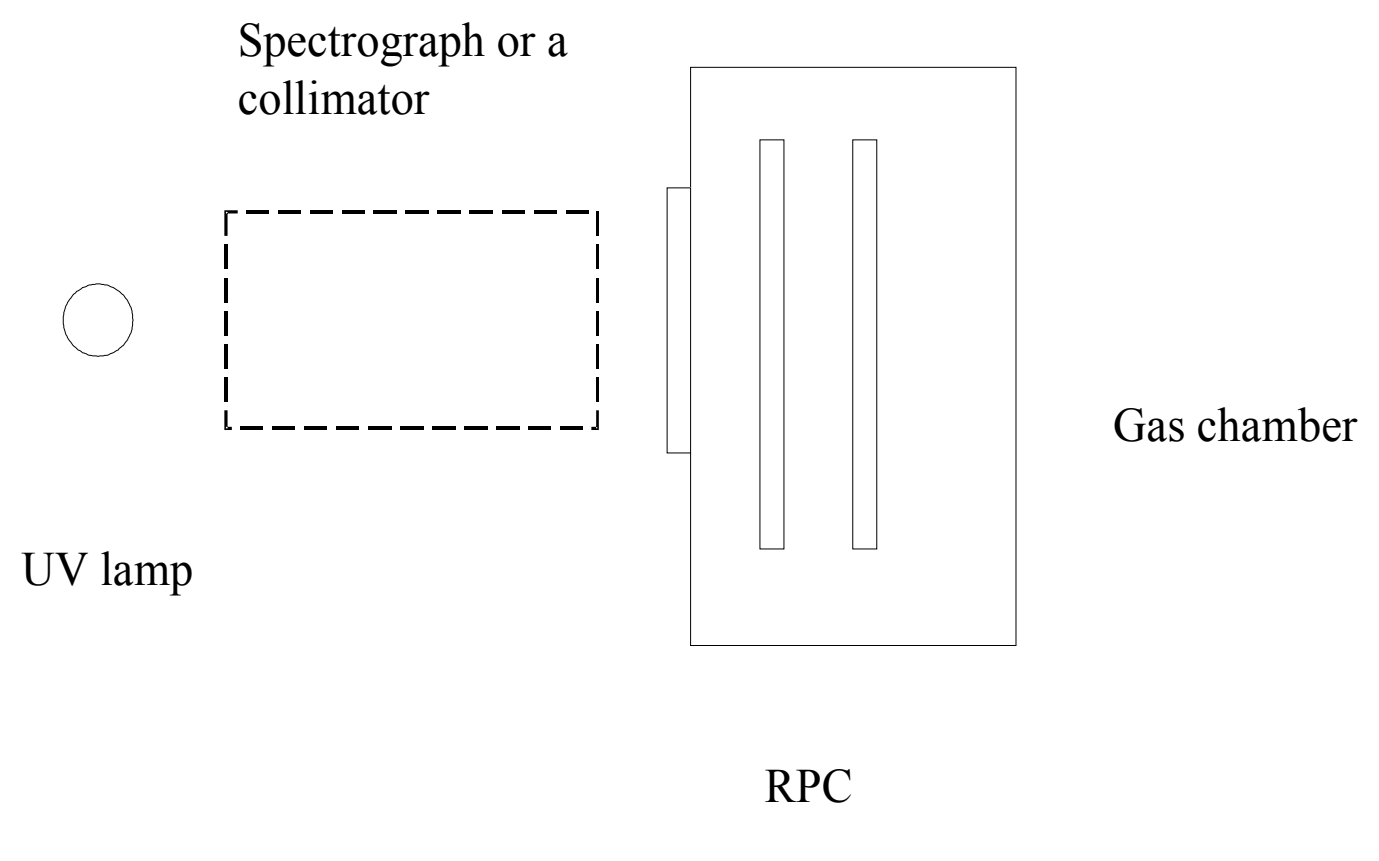

Fig1. A schematic drawing of the experimental set up

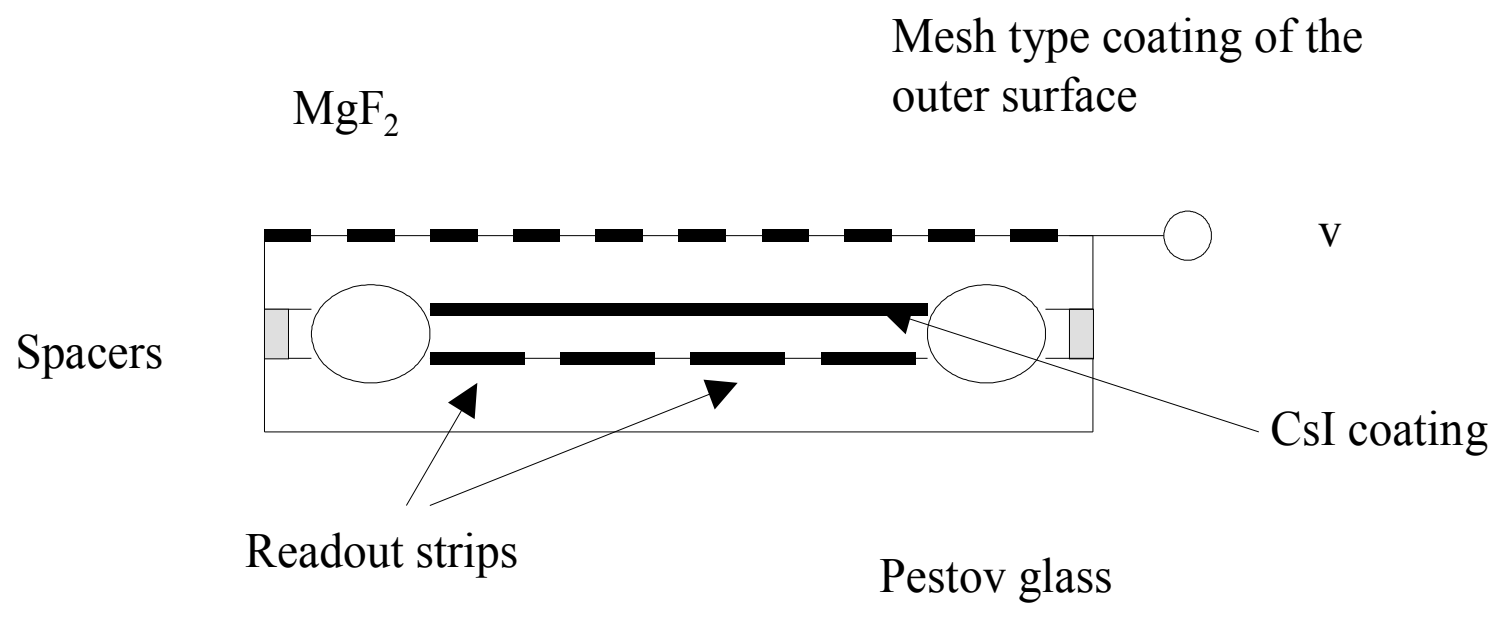

Fig.2. A schematic drawing of the RPC

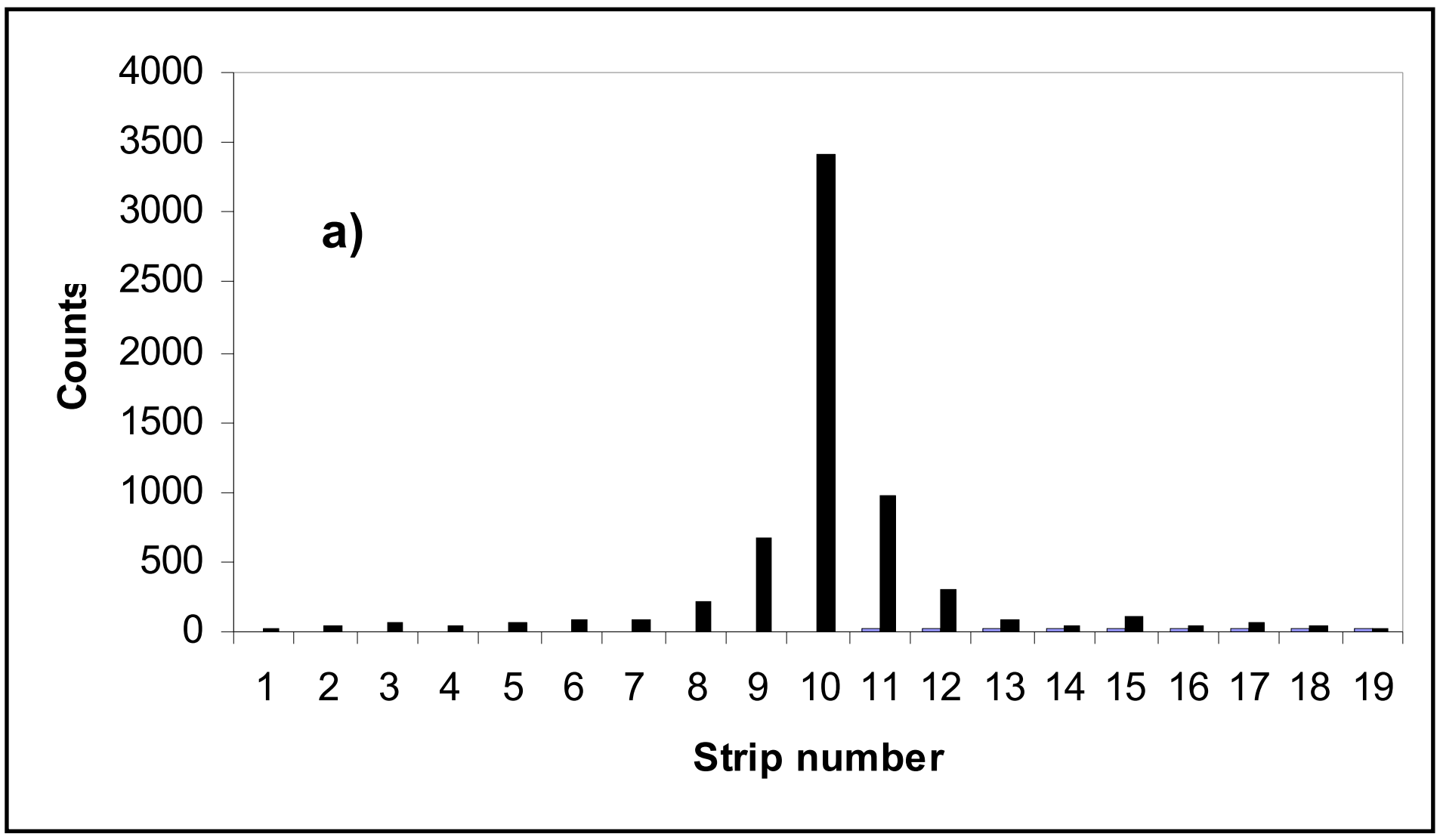

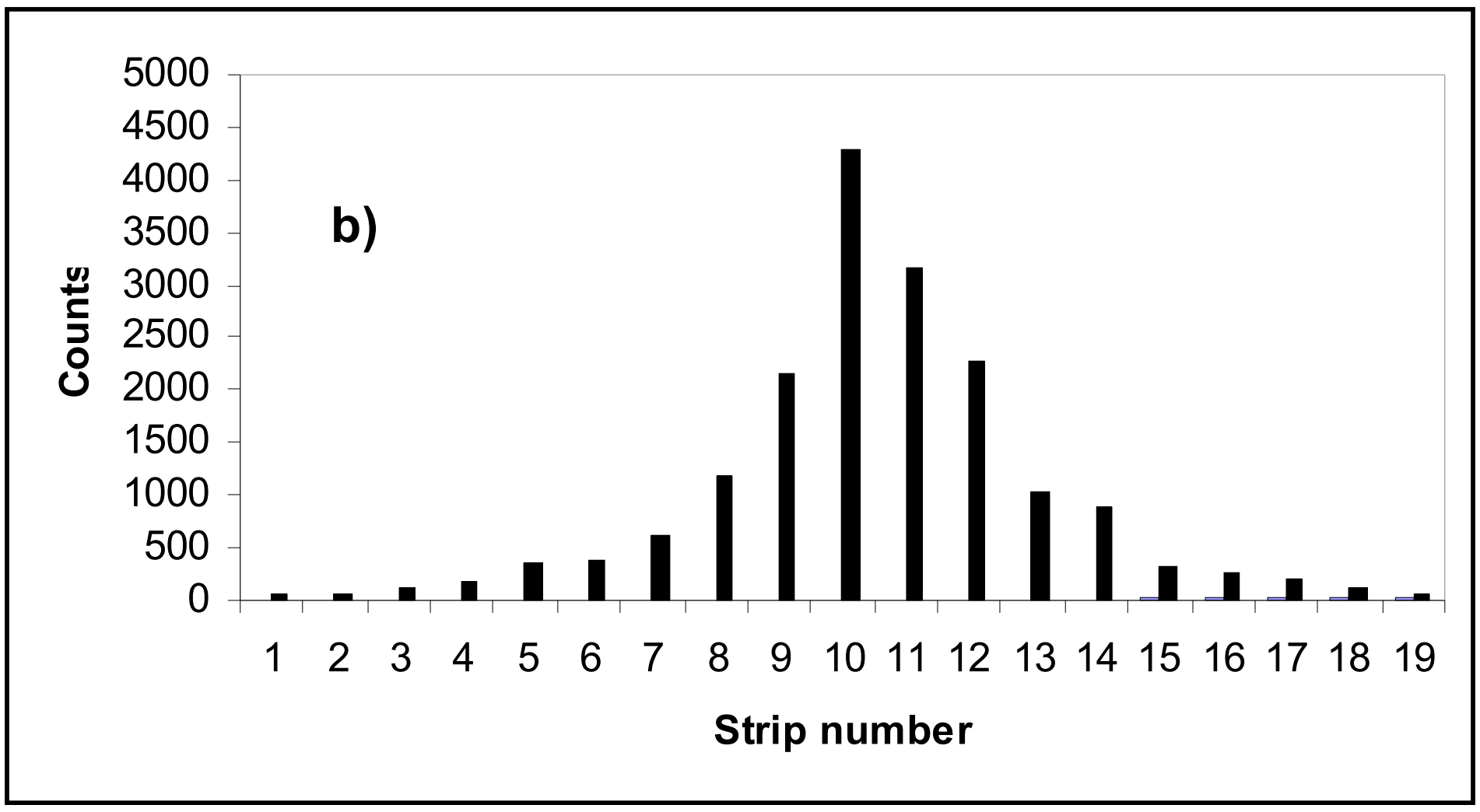

Fig. 3. Counting rate vs. strip number measured with a collimated UV beam (30 μm in diameter) in He+0,8%$CH_4$+EF mixture at gain A=of 8x$10^4$ (a) and A~$10^6$(b)

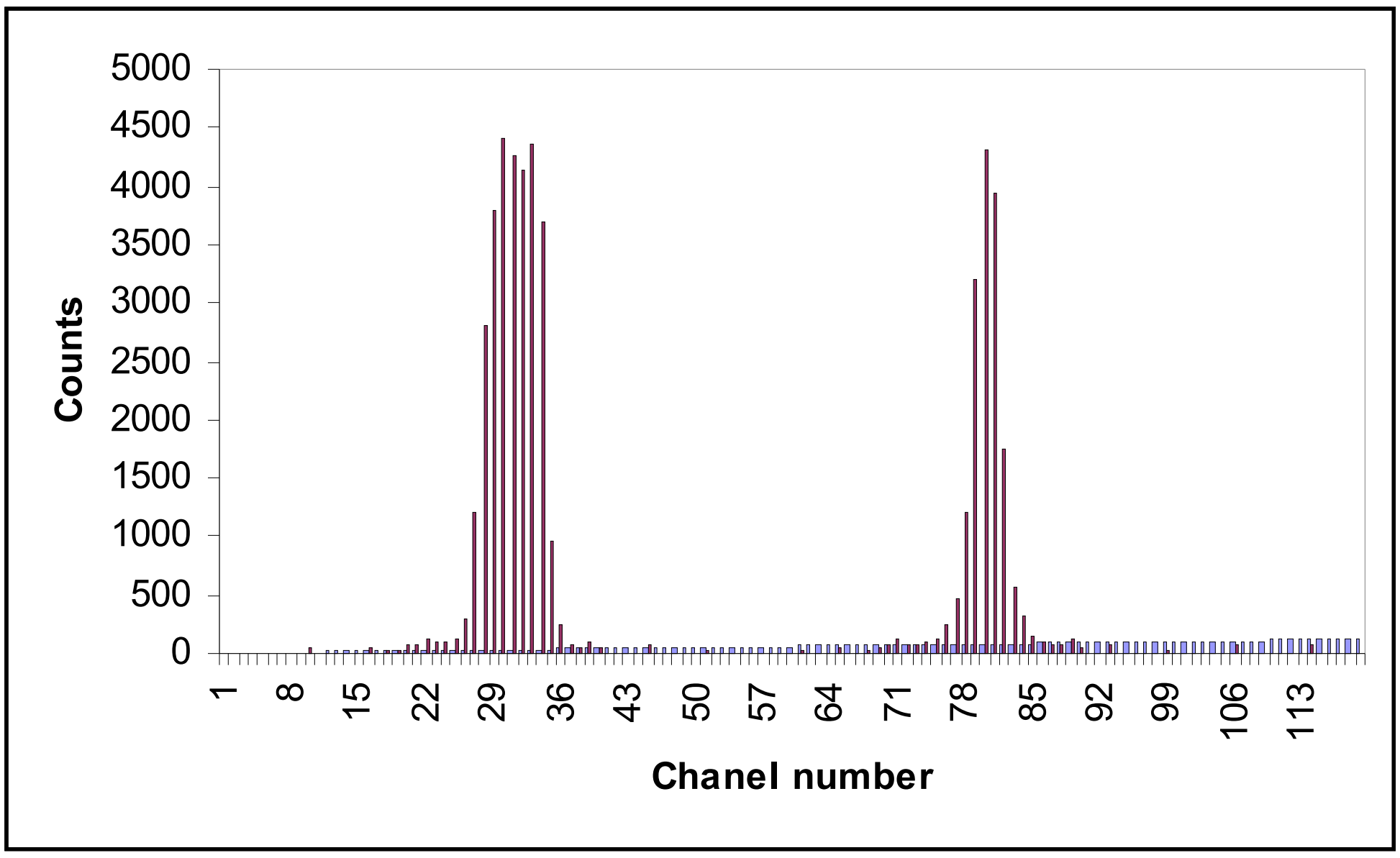

Fig.4. Images of two micro slits 250 μm and 150 μm width measured at gain of ~$10^5$.

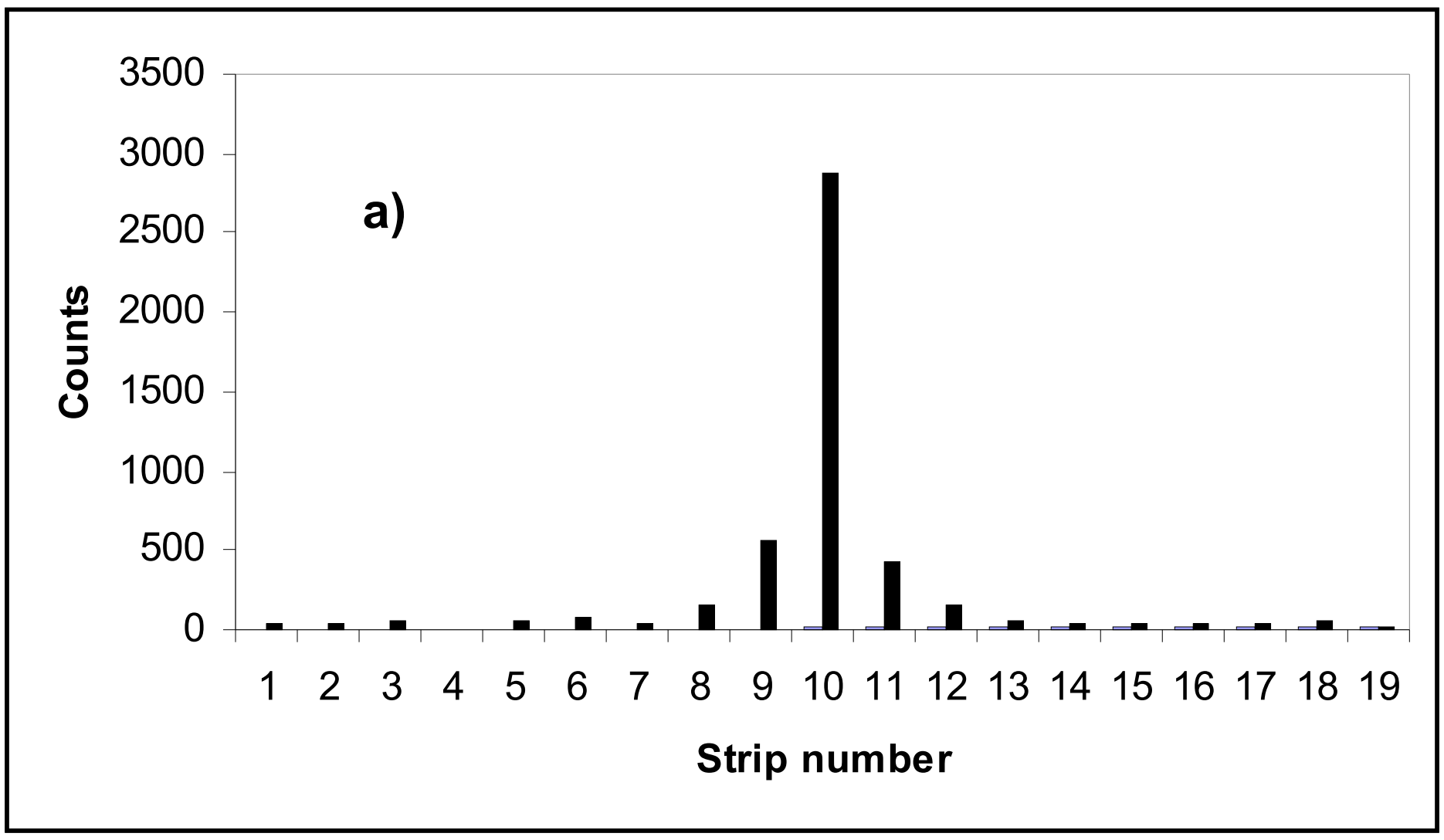

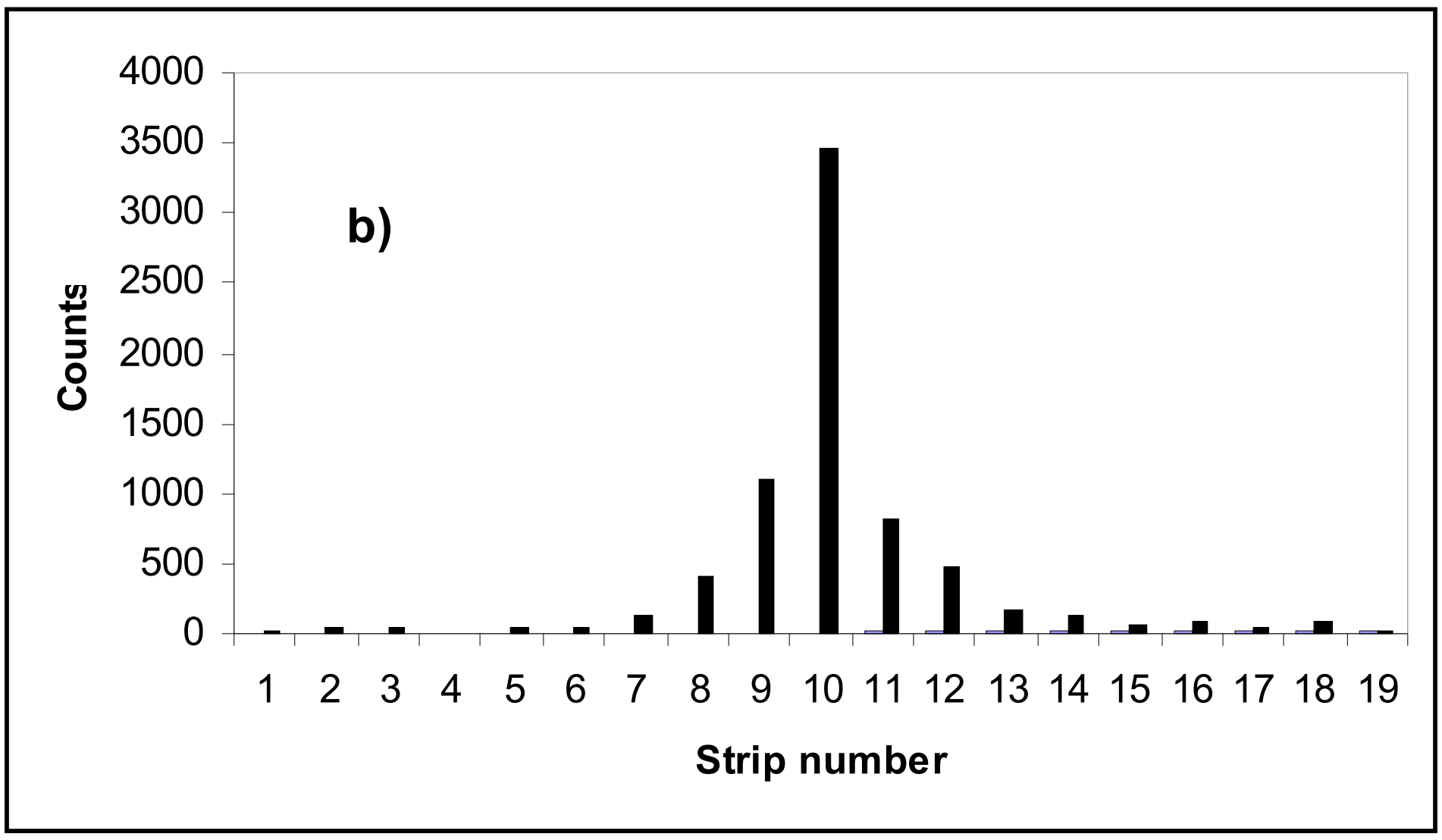

Fig. 5. Counting rate vs. strip number measured with a collimated UV beam (30 μm in diameter) in $C_2H_2F_4$+10%$SF_6$+5% isobutene mixture at gain A=of 8x$10^4$ (a) and A=8$10^5$(b)

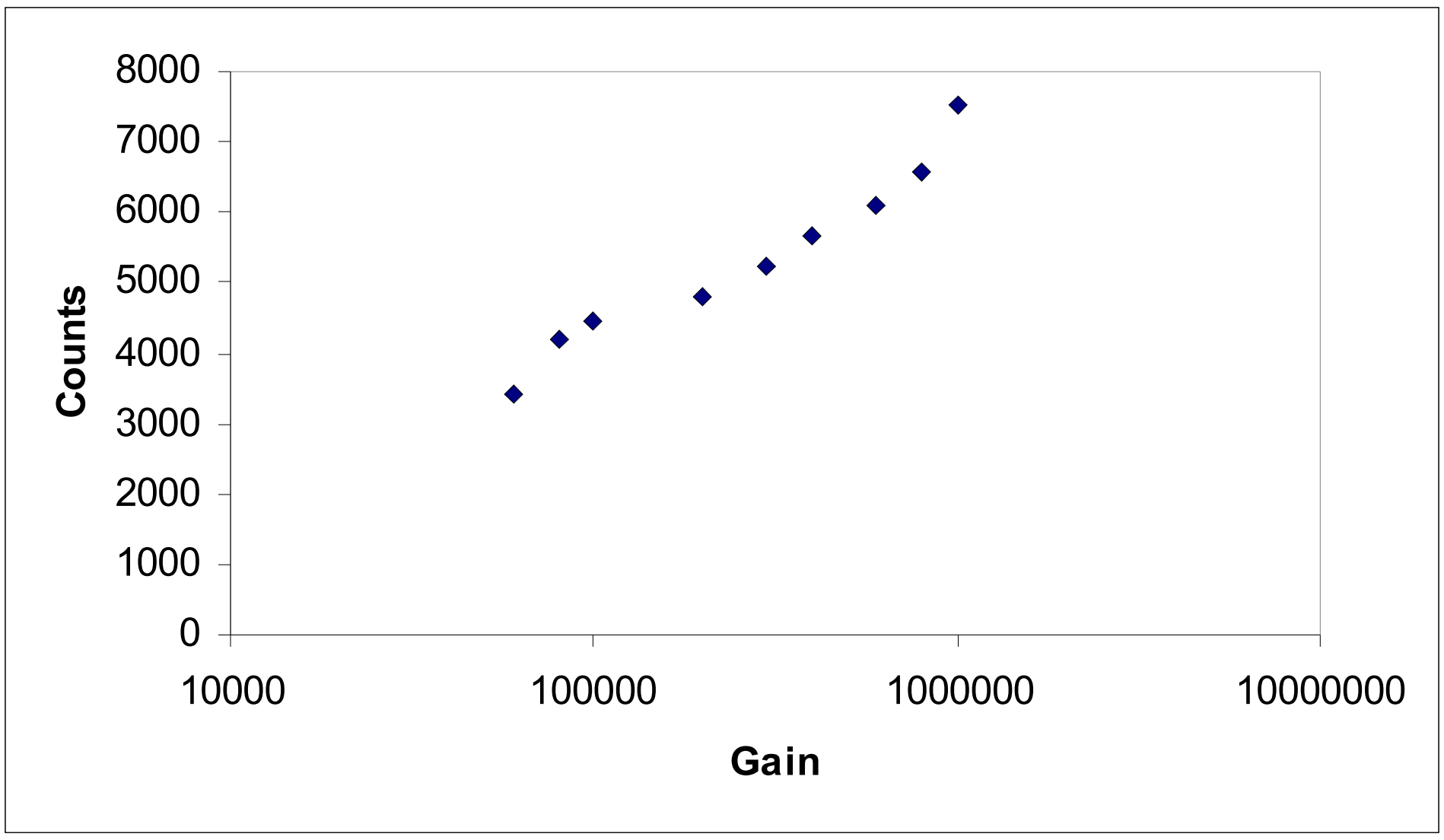

Fig.6 Counting rate vs. gain in $C_2H_2F_4$+10%$SF_6$+5% isobutene gas mixture

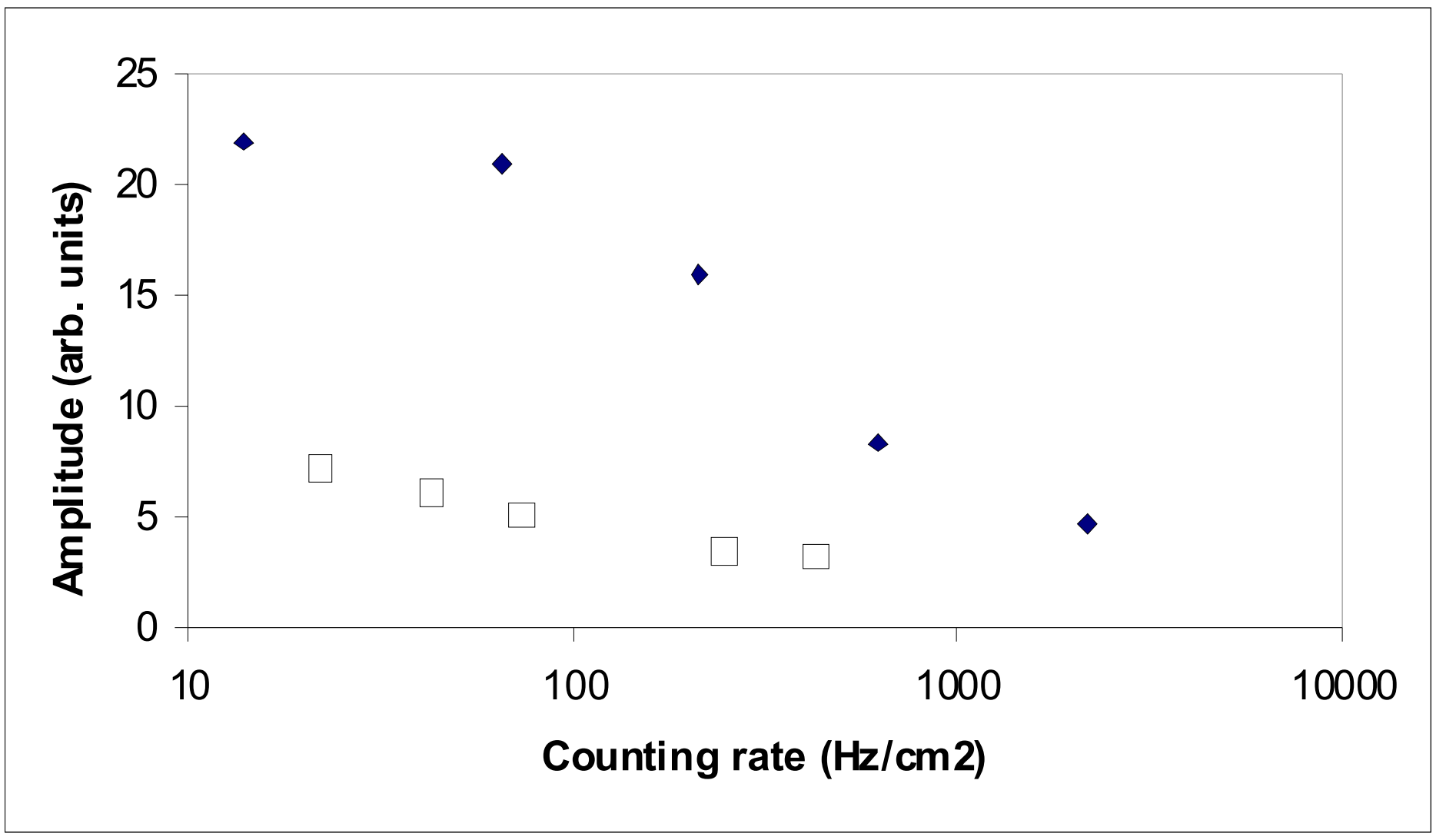

Fig.7. Signal amplitude vs. rate for two gas gains

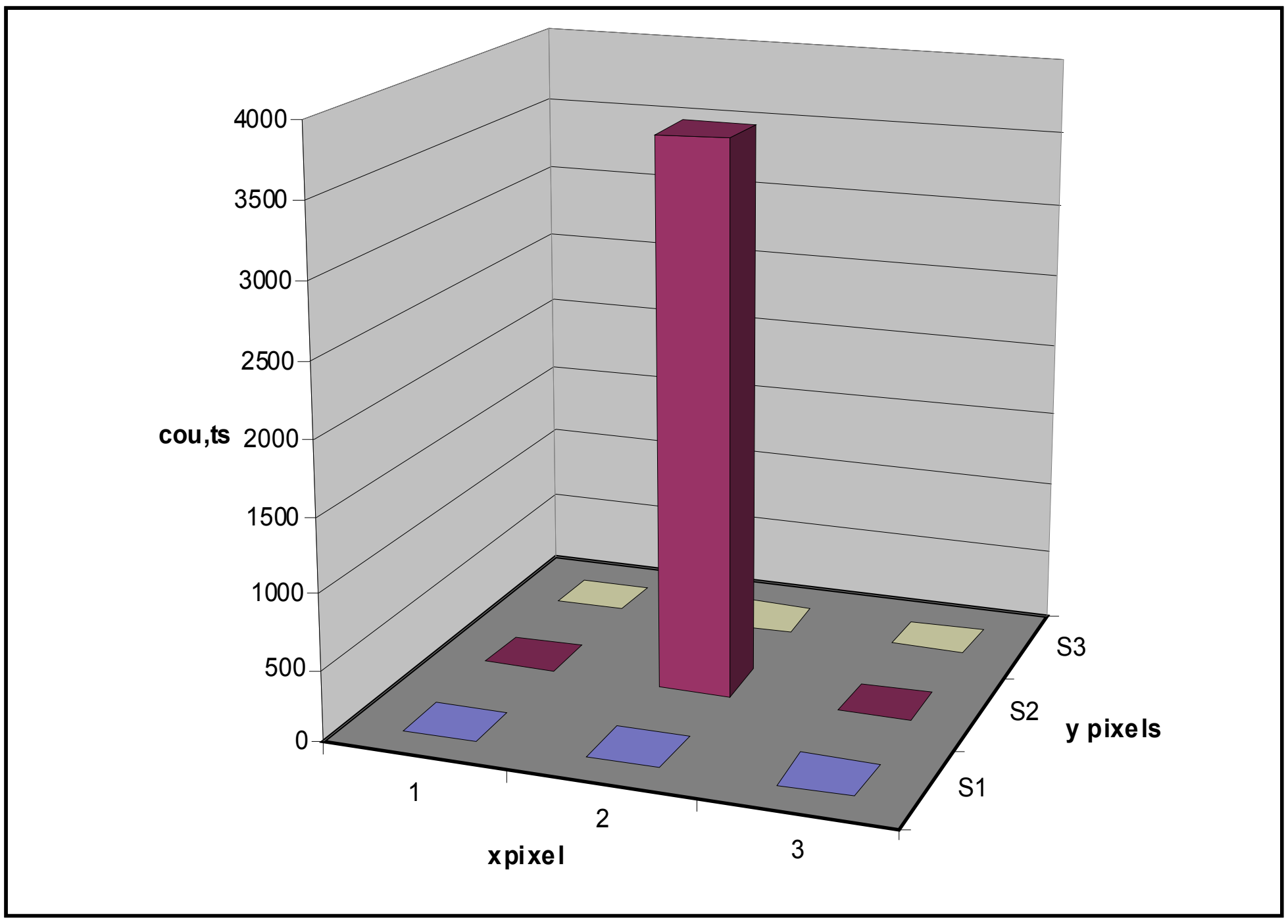

Fig. 8. Digital image if the hole 100 in diameter obtained with a pixelilized G10 anode plate

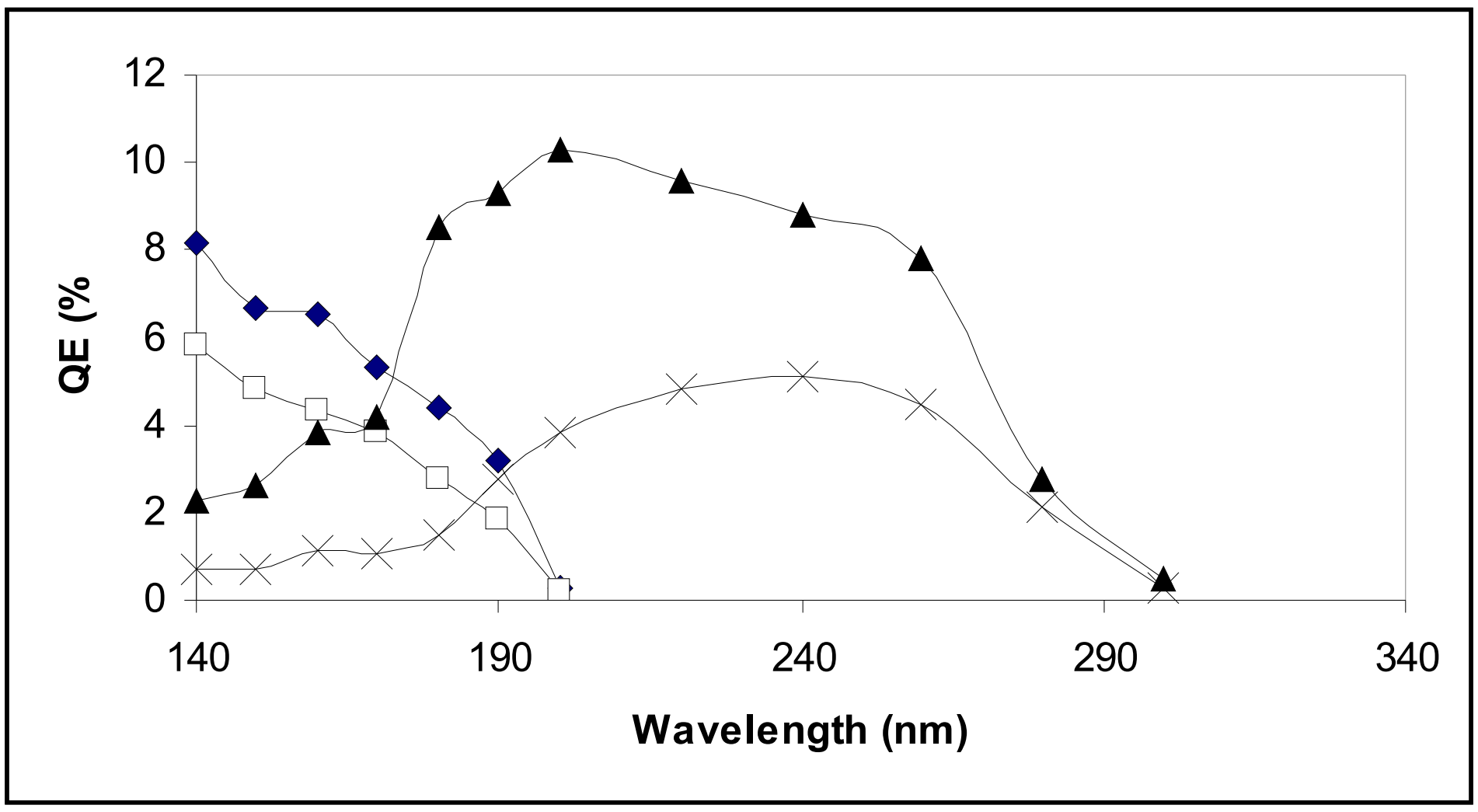

Fig.9 . Quantum efficiency vs. for various detectors: rhombus-RPC, He+0.8%$CH_4$+EF; open squares-RPC, $C_2H_2F_4$+10%$SF_6$+5%isobutene; triangles- vacuum photodiodes , starts-photodiode filled with He+0.8%$CH_4$+EF

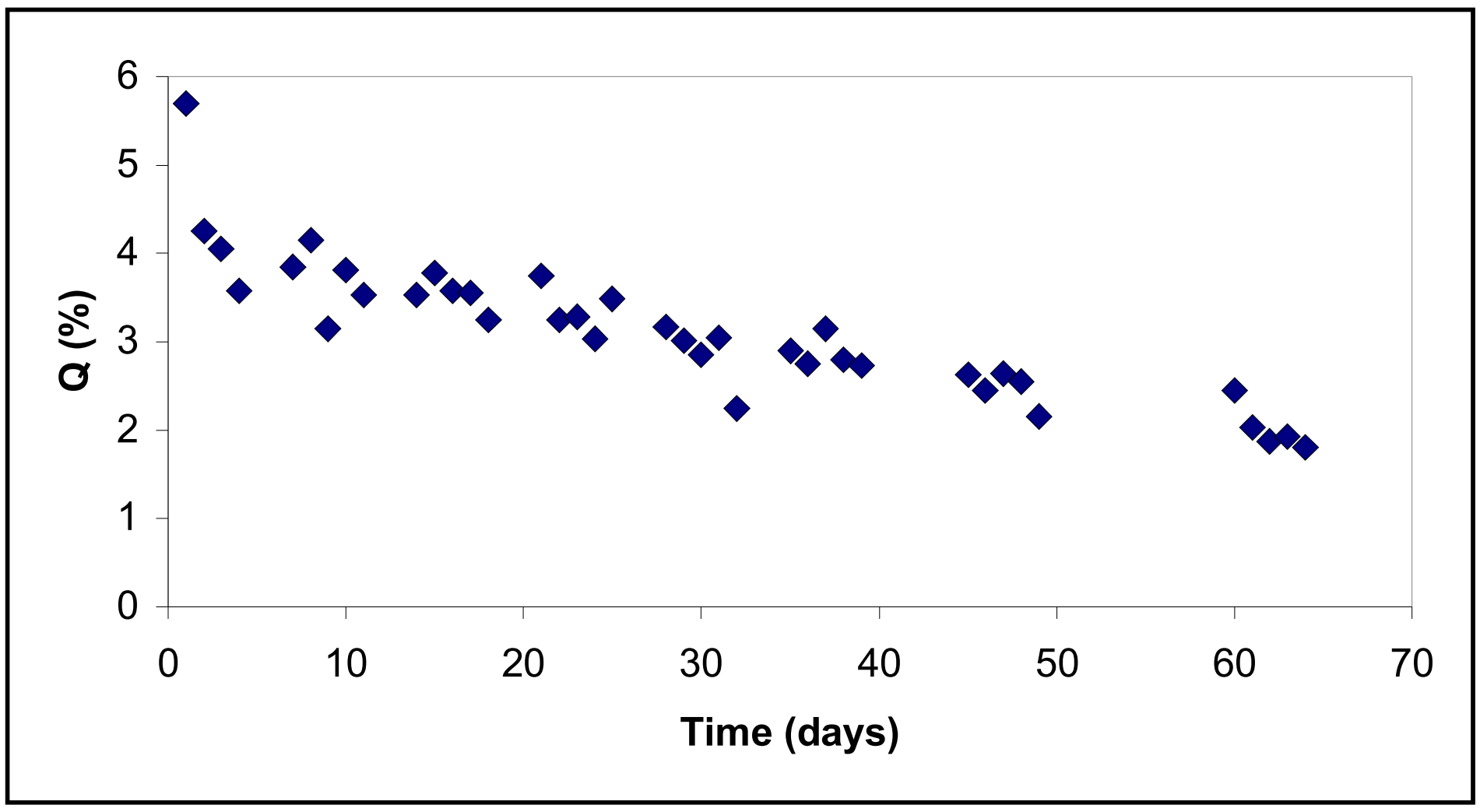

Fig. 10. Changes of the quantum efficiency with. time for a sealed detector with a CsTe photocathode. Gas mixture Ar+10%$CH_4$.

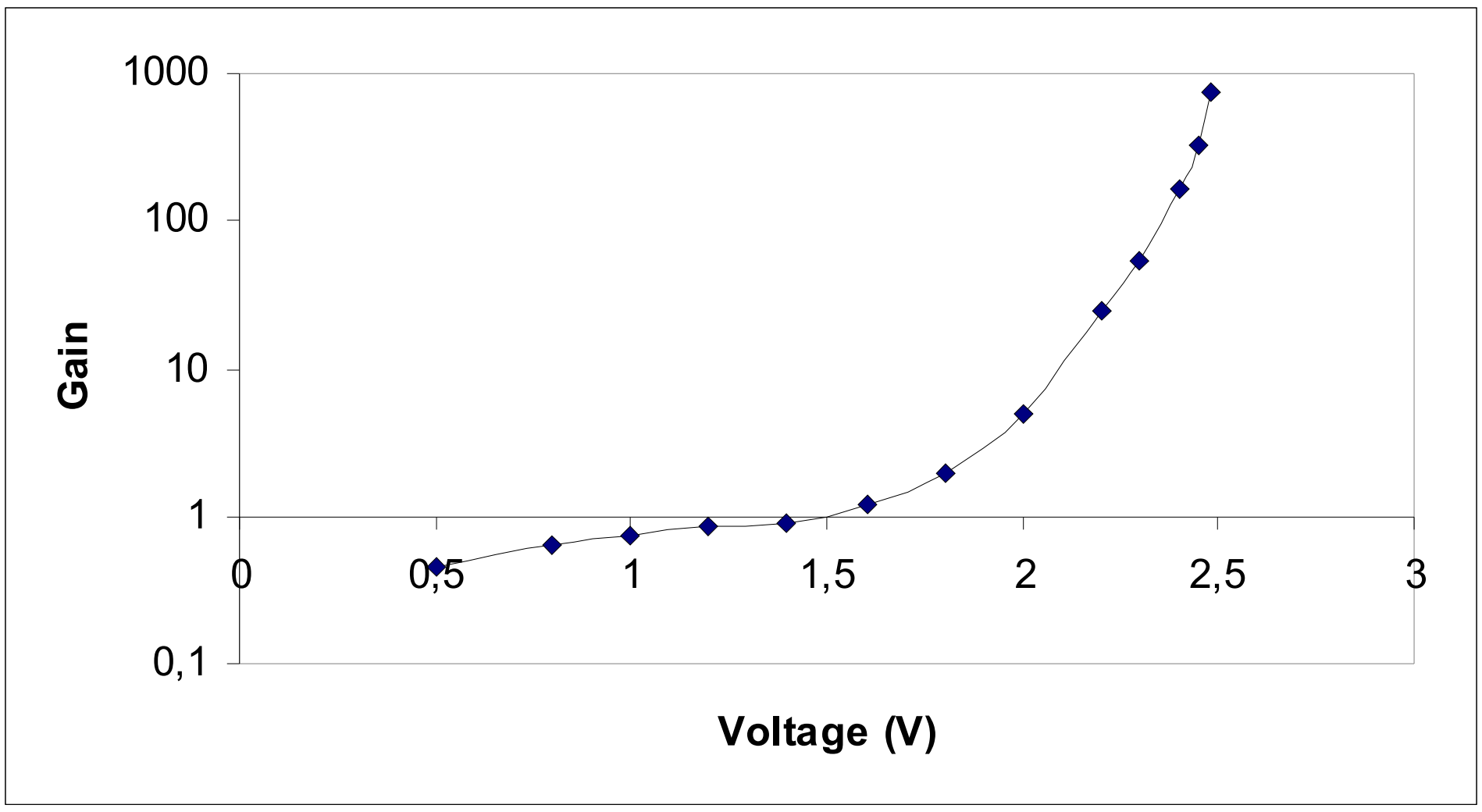

Fig. 11. Gain vs. voltage for gas –filled photodiode (He+0.8%$CH_4$+EF gas mixture)

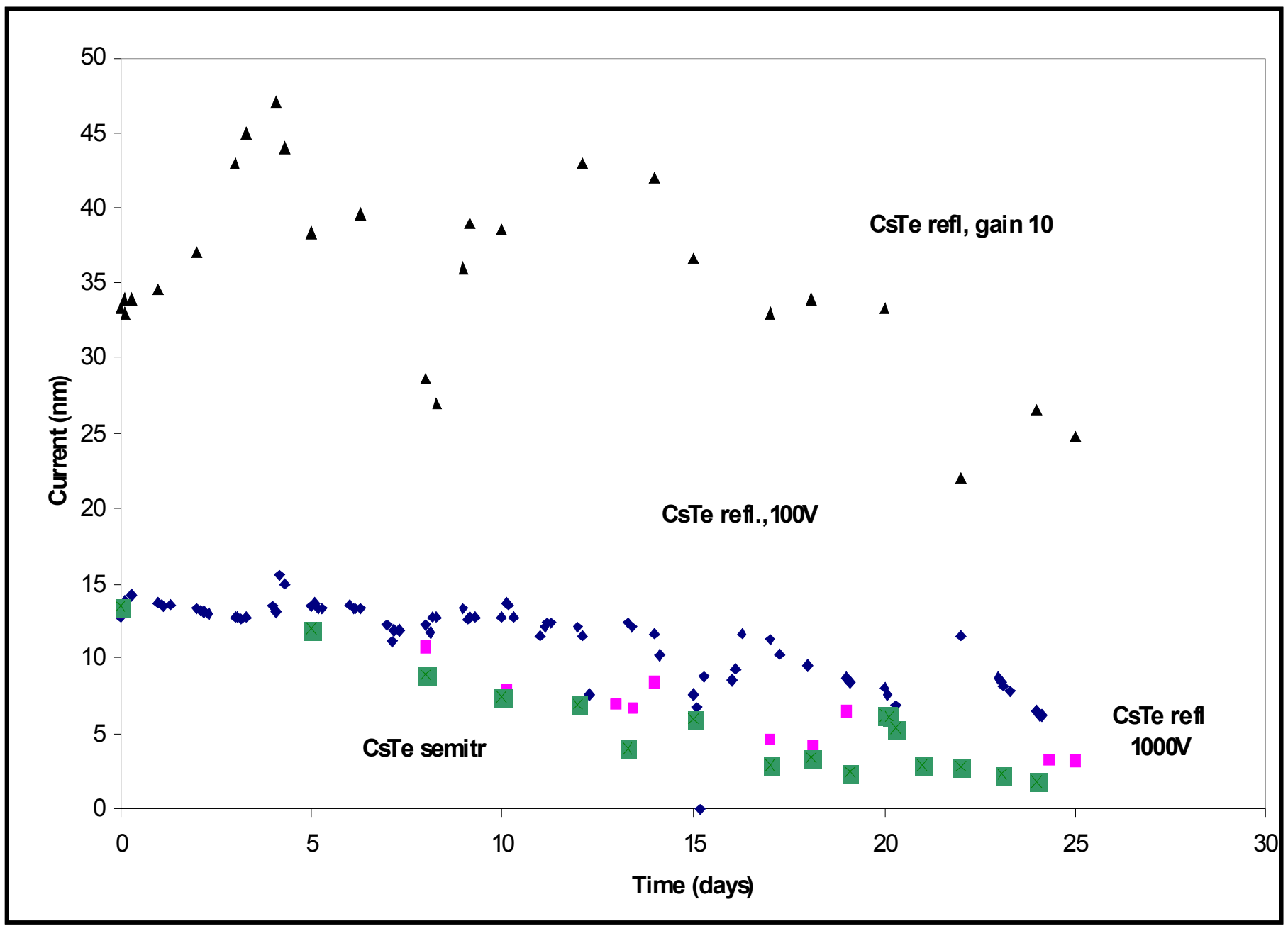

Fig.12 Aging measurements for vacuum (at applied voltages of 100 and 1000V ) and gas-filled photodiodes (Ar+$CH_4$) with CsTe photocathodes at gain of 10.

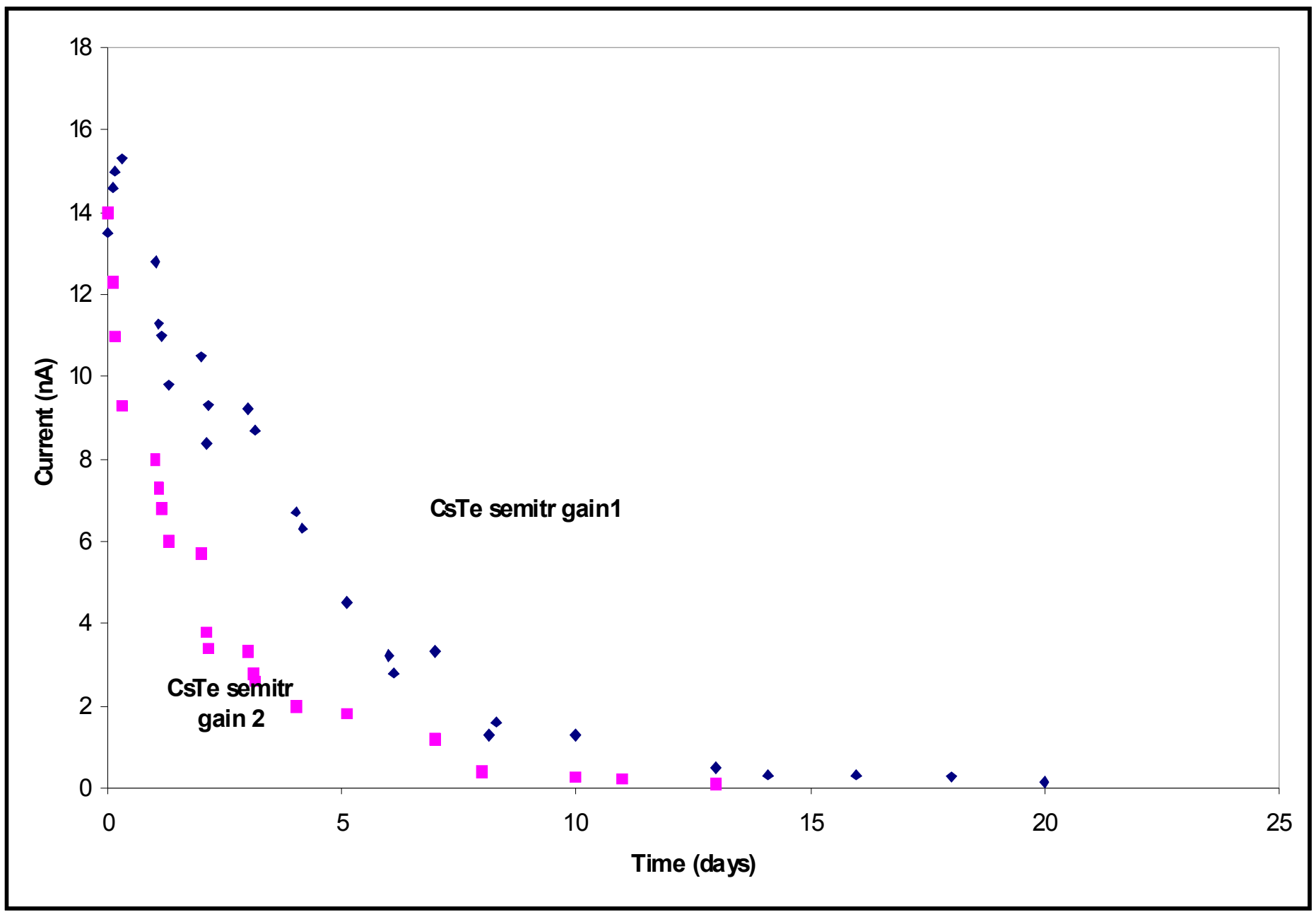

Fig.13 Results of aging measurements for photodiodes with CsTe photocathodes in Ar+isobutane gas mixture

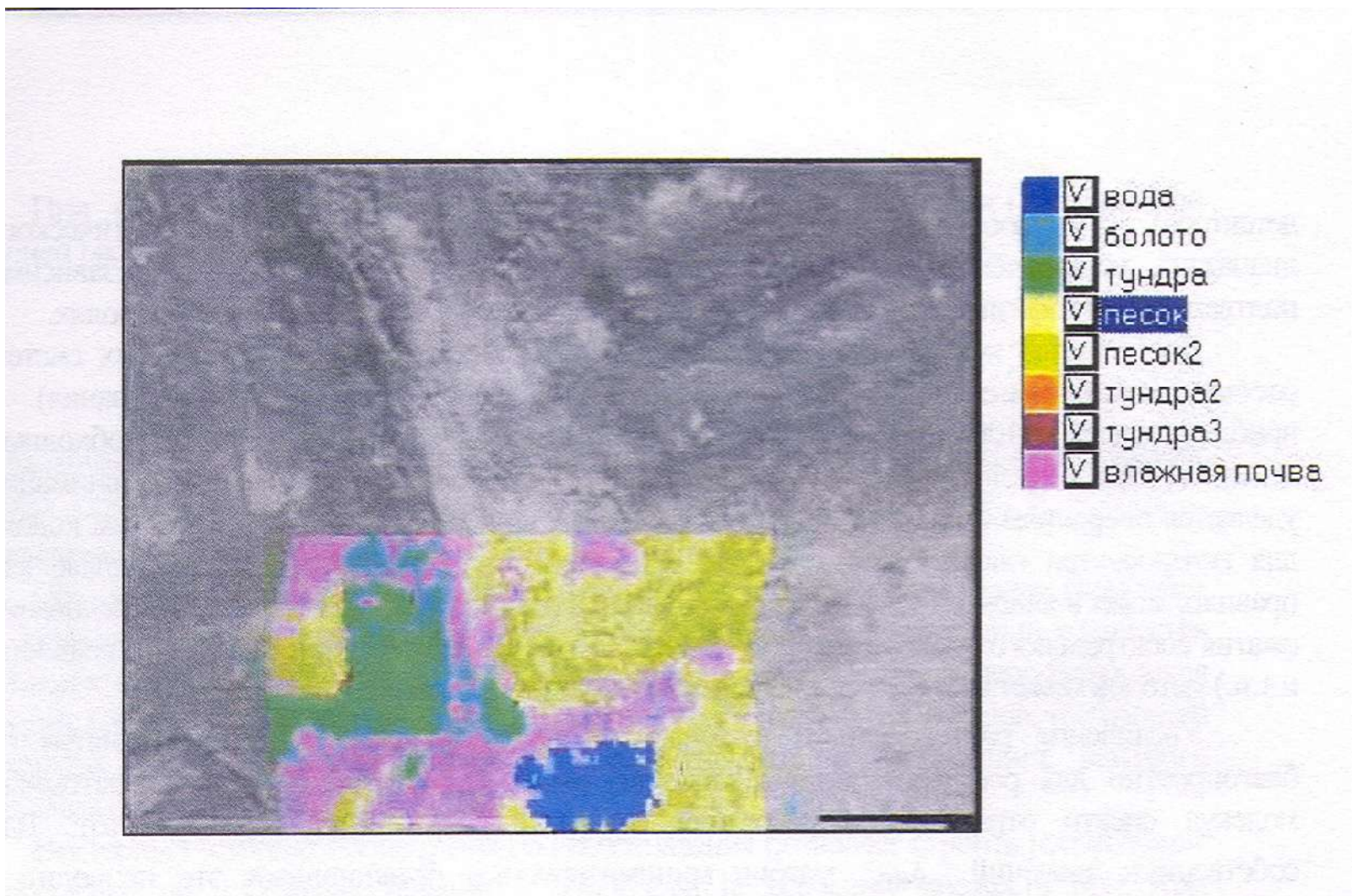

Fig. 14 Usual (gray) and hypersectrographical (color) images of earth surface [24].

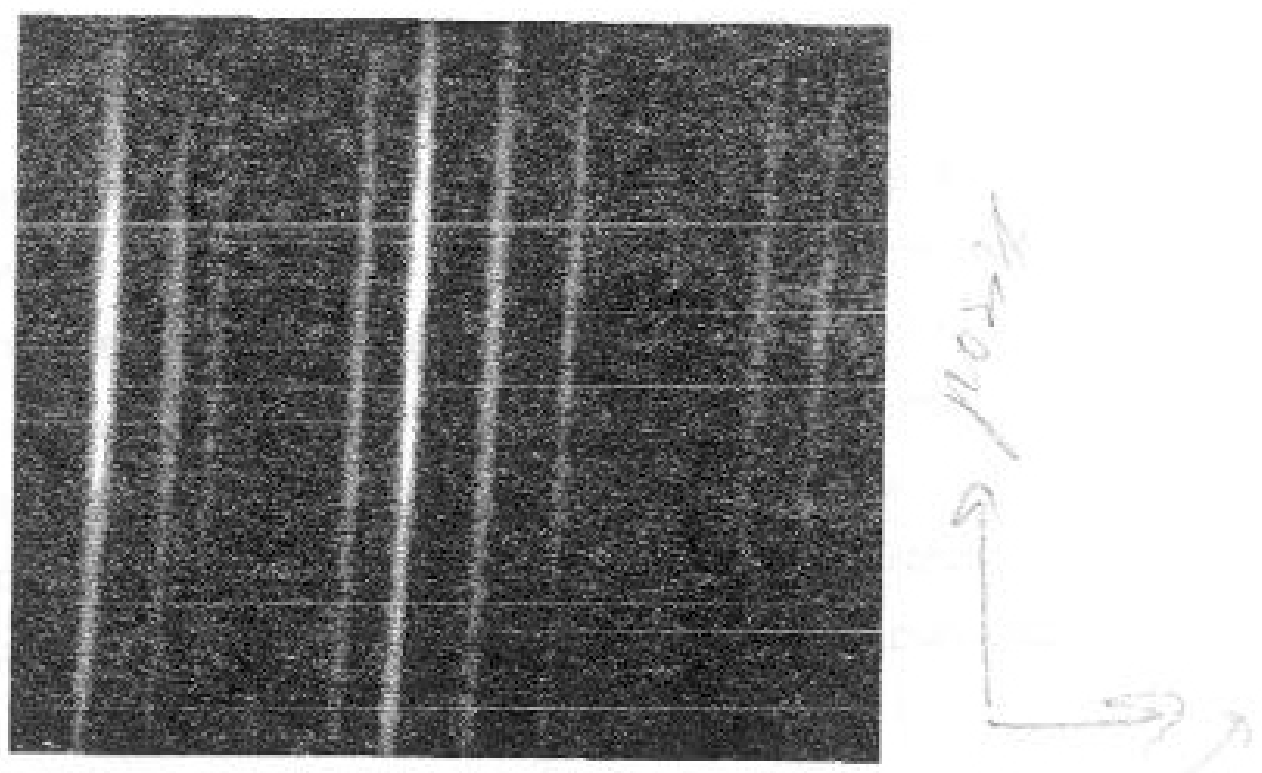

Рис.5. Изображение кадра, полученного в лабораторных условиях при освещении помещения люминесцентными лампами.

Fig.15 Hyperspectrographical UV image of the surface [24] obtained with advanced MCPs [22]. In this image the X-coordinate represents the spectra ( λ) and , Y coordinate the position along the surface

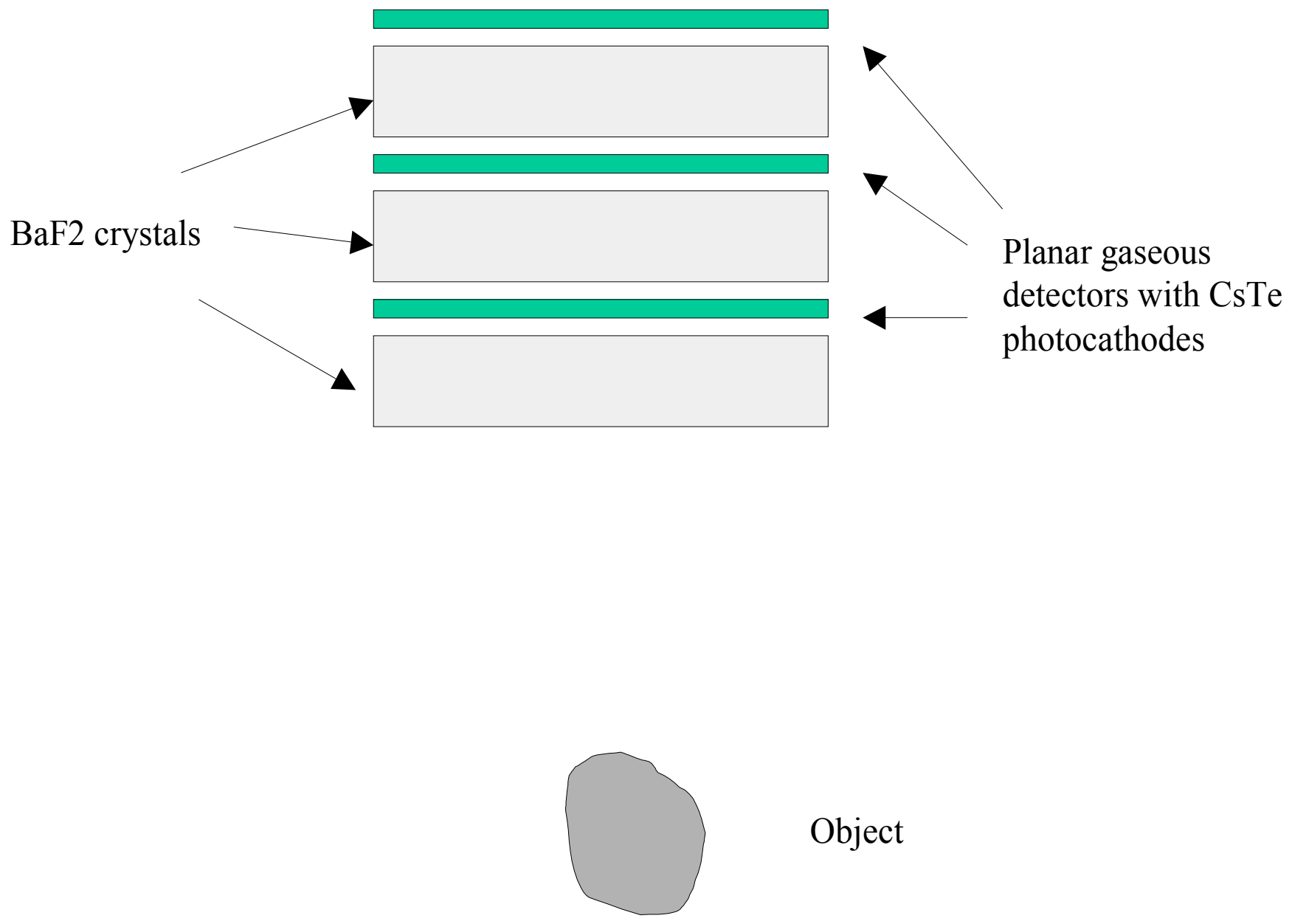

Fig. 16 The schematic view of the multiplayer time of flight $BaF_2$ PET module

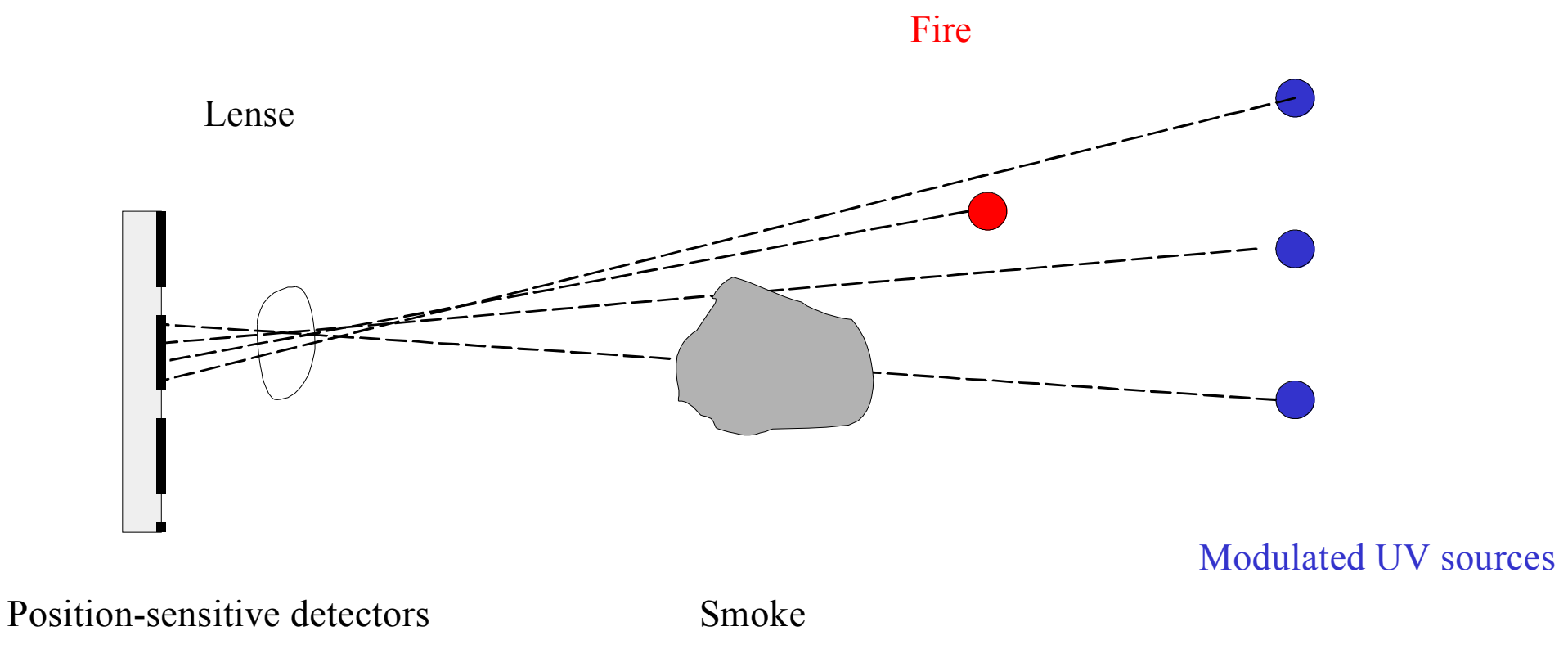

Fig. 17. A schematic drawing of the multifunctional position sensitive flame and smoke detector

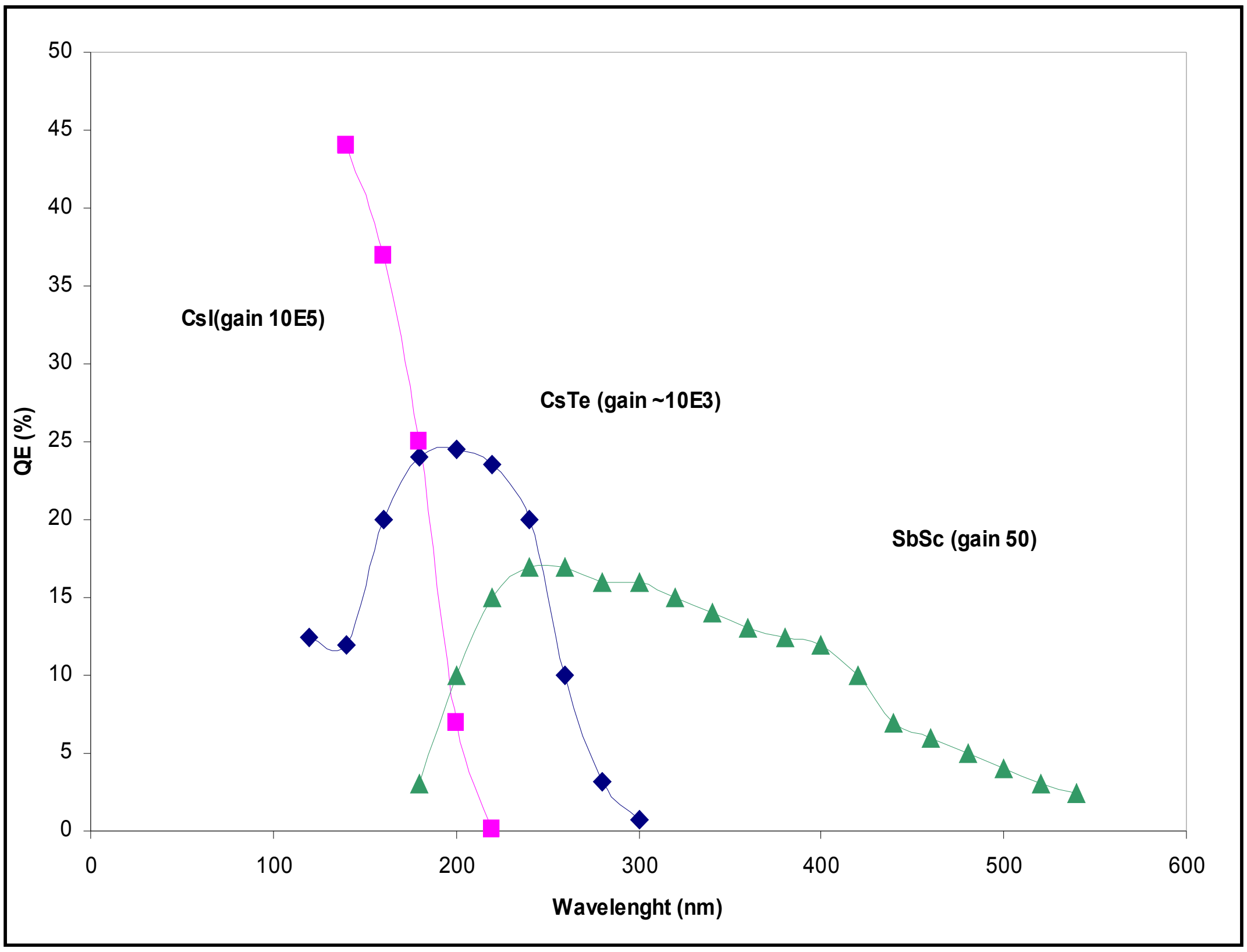

Fig 18. Quantum efficiency of CsI, CsTe and SbSc photocathodes and gains achieved in parallel-plate gaseous detectors combined with these photocathodes